\documentstyle[epsfig]{aipproc}
\newcommand{\dmm}{\mbox{$\Delta$m$_{15}(B)$}}
\newcommand{\kms}{\mbox{km s$^{-1}$}}

\newcommand{\solarm}{\mbox{$M_\odot$}}
\newcommand{\teff}{\mbox{$\rm{T}_{eff}$}}

\begin{document}
\title{The Observations of Type Ia Supernovae}

\author{Nicholas B. Suntzeff$^*$}
\address{$^*$Cerro Tololo Inter-American Observatory, 
NOAO\thanks{NOAO is
operated under cooperative agreement with 
the National Science Foundation.}\\
Casilla 603, La Serena, Chile\\}

\maketitle

\begin{abstract}

The past ten years have seen a tremendous increase in the number of Type Ia
supernovae discovered and in the quality of the basic data presented. The
cosmological results based on distances to Type Ia events have been
spectacular, leading to statistically accurate values of the Hubble constant
and $\Omega_{M}$ and $\Omega_\Lambda$.

However, in spite of the recent advances, a number of mysteries continue to
remain in our understanding of these events. In this short review, I will
concentrate on unresolved problems and curious correlations in the data on
Type Ia SNe, whose resolution may lead to a deeper understanding of the
physical mechanism of the Type Ia supernova explosions.

\end{abstract}

\section*{Introduction}

The study of supernovae is certainly in vogue today. Type II plateau and Type
Ia supernovae are used to measure distances to galaxies into the quiet Hubble
flow out past 10,000 km s$^{-1}$ to determine the Hubble constant (see for
instance \cite{Sunetal99,Jhaetal99,Gibetal00} for Type Ia distance scales, and
\cite{Schetal94} for Type II Baade-Wesselink distances). Type Ib/c in a number
of cases, and possibly IIn supernovae are associated with gamma-ray bursts and
it is now argued that a small percentage or even a majority of GRB's are
associated with core-collapse supernovae with small or non-existent hydrogen
envelopes.

Distances to Type Ia supernovae are claimed to be as accurate as 7\%, based on
the dispersion in the quiet Hubble flow (\cite{Hametal96,Rieetal96}). With such
accurate distances, a deviation from the linear Hubble flow can be measured at
higher redshifts which is related to the (de)acceleration of the local
Universe, For a precision of $\Delta m$ in distance modulus at redshift $z$,
the deceleration can be measured to an accuracy of $\Delta q_0\approx 0.9
\Delta m/z$. Statistical accuracies of 0\fm1 in an ensemble of supernovae at
$z \approx 0.5$ can lead to a error in $q_0$ of less than 0.2 units, which is
adequate to determine the sign of the acceleration.

Spectacular, and probably even believable, claims have been made about the
detection of an acceleration away from the linear local Hubble flow, which is
consistent with a flat universe ( $\Omega_{M} + \Omega_{\Lambda} = 1$) and a
positive cosmological constant $\Omega_{\Lambda}$
(\cite{Rieetal98,Garetal98,Peretal99}). It is my own peculiar viewpoint that
these results have not experienced the level of withering attacks, such as
those directed towards \cite{LauPos92}, only because the results agree with
the expectations of theory which almost unanimously require a flat universe.
Some important criticisms have been made to the interpretation of the results
(\cite{Hofetal99,Dreetal99,Agu99a,Agu99b}) but the small number of papers in
response to the results from the Supernova Cosmology Project (Saul Perlmutter,
PI) and the High-Z Supernova Search Team (Brian Schmidt, PI) lead me to
believe that these papers have not been carefully read by those who use the
results in support of their own Universe.  As a co-founder of the High-Z
Supernova Search Team, I do believe our results (see the excellent summary by
Bob Kirshner in this volume), but there are still some puzzling effects in the
data, as well as the rather loose arguments on dust and evolution that we need
to continue to perfect. Craig Wheeler, in his summary notes to this
conference, has eloquently expressed some of the weak points of the use of
Type Ia supernovae at high redshifts.

Few people question the use of Type Ia supernovae as standard candles,
corrected to a standard luminosity by some modified form of the Phillips
relationship (\cite{Phi93}) corrected for reddening
(\cite{Rieetal96,Phietal99}). The empirical evidence from the very small
scatter in the Hubble diagram convinces us that we can use Type Ia supernovae
as standard candles. But we must remember that for these standards candles, we
are unsure as to what the progenitors are, we are still not clear on where the
explosion happens in the white dwarf nor how the flame propagates (but see
article by Hillebrandt in this conference for the progress in understanding
the physics of the flame), we do not understand the effects of age or
abundance on the explosion, and we do not know the underlying physical reason
for the Phillips relationship.

In many ways, the interpretation of what we observe in Type Ia supernovae is
more complicated than in Type II events. In Type II supernovae, there is a
nice, optically thick hydrogen envelope which hides lots of physics around the
time of maximum light, and the resulting observable ``uvoir'' or
ultraviolet+optical+infrared radiation is the scattered and thermalized gamma
rays during the transfer through the photosphere. For Type Ia supernovae, the
uvoir optical depths in mass are much larger, and in some wavelength regions,
the ejecta are transparent at maximum light. The mass encountered by an
escaping gamma ray is smaller in the Type Ia supernovae compared to Type II ,
and the gamma rays leak out in ever larger quantities near maximum light
(\cite{LeiPin92} and this volume.) This also causes the local ionization and
excitation equilibria to be affected by more distant parts of the ejecta
leading to much more complicated NLTE effects than in Type II supernovae near
maximum light.

One important simplifying condition in a Type Ia supernova is that the ejecta
must expand from the size of a white dwarf to the size of a star before it is
bright enough for us to observe it (via $L \propto T^4R^2$ with $T$ about
10000K). This expansion takes about 20 days to optical maximum (see the
article by Aldering in this volume), and with such a long expansion time, many
of the detailed effects on the explosion from different progenitor models get
washed away over the large expansion. This is partly why the lack of an
agreed-upon progenitor model is less embarrassing than it otherwise would be.

Rather than review the properties of Type Ia supernovae for this conference, I
would rather discuss a few areas of topical interest in the observations of
Type Ia's. These are areas in which I think there will be progress in the next
five years, and the resolution of some of these topics will help us in
understanding the physics of the explosions and ultimately the utility of Type
Ia supernovae as standard candles.

Those seeking a review of supernovae, and of Type Ia supernovae in particular,
should consider the following resources. An recent excellent and exhaustive
book discussing supernovae from a theoretical perspective has been written by
Arnett (\cite{Arn96}). Another excellent theoretical perspective is given by
Wheeler et al.(\cite{Wheetal95}). The physics of the explosion is a rapidly
evolving subject and the most up-to-date review is \cite{HilNie00}. Recent
books covering the general field of supernovae from the perspective of invited
authors are: \cite{McCWan96,Ruietal97,NieTru99}. The spectral properties of
supernovae are given by Filippenko (\cite{Fil97}). An excellent general review
of the optical properties of Type Ia supernovae has been written by Leibundgut
(\cite{Lei00}).  Unfortunately, there is no recent monograph on the photometry
of Type Ia supernovae to my knowledge.

\section{Topics of Interest in Type Ia Supernovae}

\subsection{Age, Metallicity, and Reddening}

The basic observational result of the Type Ia supernova searches at $z \approx
0.5$ is that the corrected peak brightness is fainter by about 0\fm25
magnitude than predicted by a cosmological model with $\Omega_M=0.2$ and
$\Omega_\Lambda=0$. Adding matter to the universe only makes the predicted
magnitude brighter.  What can cause this? The three suggestions are:
evolution, cosmology, or reddening.

In terms of evolution, the look-back time to this redshift is $\sim5$Gyrs.
Specific predictions about the effects of evolution have been discussed by
\cite{Hofetal99}. They consider the age and metallicity difference and how it
may affect the peak brightness of Type Ia supernovae. However, the look-back
time of 5Gyrs is a rather modest look-back in terms of the evolution of
galaxies. The evolution in metallicity in a typical galaxy is very small over
this time - after all, this is less than the age of the sun. For instance, the
metallicity evolution for our Galaxy over this look-back time was only
$\delta{\rm[Fe/H]} \sim -0.2$. This is even {\it smaller} than the natural
real scatter in the metallicity of $\sigma{\rm[Fe/H]} \sim 0.25$ among stars
at a given age in the Galactic disk  (\cite{Evaetal93}). Metallicity evolution
in the $z \sim 0.5$ sample is probably not an issue.

The issue of age is less clear. As discussed by \cite{Hofetal99}, the C/O
ratio of the soon-to-explode white dwarf is a function of age, and since the
fuel for explosion comes the carbon and oxygen, the peak luminosity is a
function of age. They find that a decreasing C/O ratio implies a brighter peak
luminosity for a Type Ia. In an attempt to parameterize this, \cite{vonetal97}
consider the evolution of a white dwarf population a function of stellar
population age. Using the sub-Chandrasekhar mass explosion models of
\cite{WooWea94}, they are able to recover the approximate range in luminosity
seen in the Cal\'an/Tololo supernova sample (\cite{Hametal96}). The supernova
rates are higher in star-bursting galaxies which are producing white dwarfs at
higher masses which require less accretion. In fact, their simple modeling
shows that for modest increases in the space density of star-burst galaxies at
$z \sim 0.5$ which is well below any observational limits on the enhancement
of star formation at this redshift, the star-burst galaxies can produce one
half of the supernovae in that volume.

Certainly the anecdotal evidence supports this. As pointed out by
\cite{vonetal97}, the modest starburst galaxy M100 has had four supernovae
(one was a Type II) whereas the huge elliptical M87 has had none. More
striking is the case of the amorphous (and active star forming) low-mass
galaxy NGC 5253 has had two SNe in the last century, but the whole Coma
cluster which has 10,000 gas-poor $L_*$ galaxies, has had only a few supernova
the last 25 years.\footnote{It is not clear, however, how carefully we are
looking for supernovae in Coma, and certainly this cluster should be targeted
by northern hemisphere observers involved in CCD searches. The comparison of
supernovae in Coma and a nearby cluster such as Fornax would be an important
test for distance-scale calibrations.} More rigorous tests (\cite{Capetal97})
show that spiral galaxies have higher Type Ia rates per unit luminosity than
elliptical galaxies by roughly a factor of two. It has even been claimed that
Type Ia supernovae are associated with spiral {\it arms} (\cite{Baretal97}.

Given the fact that Type Ia supernovae are more likely to occur in spiral
galaxies, yet they also appear in normal elliptical galaxies, implies that both
young and old populations can produce Type Ia supernovae. Average ages for
elliptical galaxies are probably at least as old as 5Gyrs which means that the
{\it local} sample of host galaxies of supernovae probably cover the same
range in average population age as the look-back time to $z \sim 0.5$.  {\it
Thus the local supernovae used in the calibrations of the Phillips effect
probably cover the parameter range in age and metallicity implied by a look-back
time of 5 Gyrs.} In detail, it is important to check this by comparing the
slope and zero-point of the relationship for the nearby and $z \sim 0.5$
samples, and by checking whether the ellipticals and the spirals give the same
relationship locally.

The final question is that of reddening. The cosmological result depends on a
measurement of 0\fm25, which corresponds to about $E(B-V)=0.08$. Apparently
the mean colors of the nearby and distant samples are inconsistent with the
distant sample being reddened by 0.08. A novel form of dust has been invented
by \cite{Agu99a,Agu99b} which, by the way of formation, is more grey than
typical dust in the disk of a galaxy. Clearly multicolor observations of
supernovae at $z \sim 0.5$ (which will require near-infrared photometry) are
needed to sort this out. But a more mundane question arises - do we understand
the intrinsic colors of nearby supernovae?

\begin{figure}[t] 
\centerline{\epsfig{file=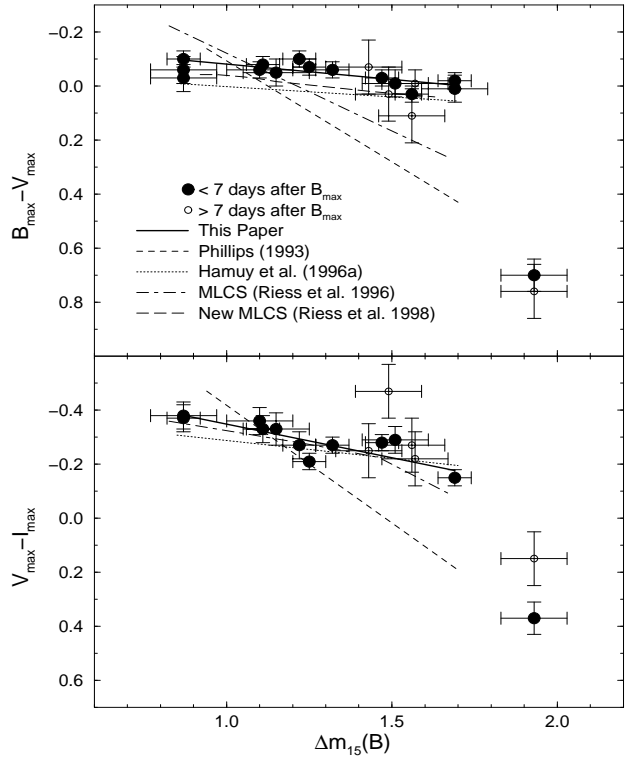,height=4.5in,width=3.5in}}
\vspace{10pt}
\caption{The ($B_{max}-V_{max}$) and ($V_{max}-I_{max}$) ``color'' is plotted
versus the decline rate parameter \dmm\ for 20 events with $E(B-V) < 0.06$.
These SNe are likely to have little or no host galaxy reddening.  Note that
these colors are not the $BVI$ colors at $B$ maximum light, but are calculated
from the maximum light magnitudes for each broad-band magnitude. For instance,
the $V_{max}$ occurs about 1.5 days after $B_{max}$.  Figure from
\protect\cite{Phietal99}.}
\label{fig1}
\end{figure}

In a recent paper (\cite{Phietal99}), we have reexamined the reddening of Type
Ia supernovae based on supernovae we suspect of having little or no reddening
due to material in the host galaxy. Choosing ``unreddened'' supernovae to
define a baseline sample is not easy. At low dispersion, the absence of
interstellar lines at the 0.1A level still does not rule out significant
reddening. Only with high S/N high-dispersion spectra, which are rare for SNe,
can one really directly verify if there is reddening. One can also chose SNe
in elliptical or S0 galaxies away from obvious dust lanes as probably
unreddened.  One must worry, however, that the properties of the
``unreddened'' sample are not biased by chosing the sample based on galaxy
type. In Figure \ref{fig1} I show our best guess for the intrinsic color range
for Type Ia supernovae as a function of the luminosity parameter
$\Delta{m}_{15}$ used by our C\'alan/Tololo group as published in
\cite{Phietal99}.

Figure \ref{fig1} shows how radically our idea of the unreddened colors of
Type Ia supernovae have changed in the last 7 years. The early results
suggested a very strong color dependence as a function of \dmm. Curiously,
this color dependence mimicked the reddening correction such that the
reddening and the assumed correction to standard luminosity were degenerate, a
fact pointed out by Sidney van den Bergh. However, it now appears that Type Ia
supernovae, except for the very underluminous events like SN1991bg, are very
similar in $B_{max}-V_{max}$ and $V_{max}-I_{max}$ color at maximum light
despite the range of over 1 magnitude in intrinsic brightness. Theory must
explain why the range in Type Ia supernovae are so ``grey'' at maximum light.

In the early years of the use of Type Ia supernovae to measure the Hubble
constant, reddening was not even considered. It was assumed that the
distribution of reddening in the local and Hubble flow sample was the
same. This can't be true in general, since the calibration sample comes from
host spiral galaxies with Cepheids (where there can be reddening) while the
field sample contains both ellipticals and spirals. It is obviously important
to correct for reddening and the more modern techniques, such as used by
\cite{Rieetal96,Rieetal98,Phietal99}, solve for reddening as part of the
least-squares technique. But if the true unreddened colors of Type Ia
supernovae remain somewhat uncertain as I believe, the Baysean assumption of
no negative reddening is not valid. 

Much observational work on the intrinsic colors of Type Ia supernovae needs to
be done. The inclusion of infrared photometry will aid greatly in the
definition of intrinsic supernova colors and reddening. At this time, there
are only a very few papers on infrared photometry of Type Ia supernovae
(\cite{Elietal81,Elietal85,Froetal87,Mei99,Krietal99}).


\subsection{Are We Doing Photometry Yet?}

\begin{figure}[t] 
\centerline{\epsfig{file=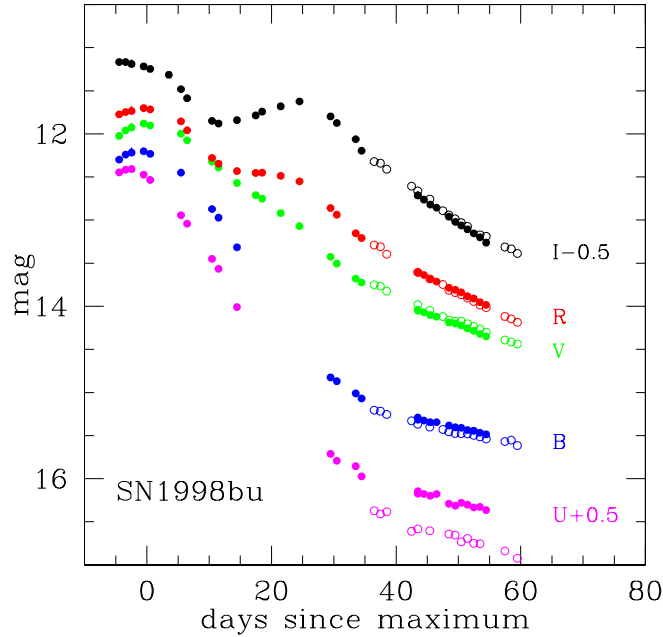,height=3.5in,width=3.5in}}
\vspace{10pt}
\caption{ The $UBVRI$ photometry for SN 1998bu. The solid circles refer to
data taken with the CTIO 0.9m telescope as published by Suntzeff, et
al. (1998). The open circles are data taken with the CTIO YALO 1m
telescope. All the data have been reduced using the same local standards.}
\label{fig2}
\end{figure}

Most people who are interested in using supernovae for cosmology will go to
the literature and read photometric numbers out of a table. If they actually
look at the paper, they will usually see some dismal discussion about local
standards, ``natural'' versus ``standard'' system, color terms, shutter
corrections, non-linearity in the detector, and the like. One is usually
reassured that the local standards repeat to some level like 0\fm02 and
therefore the photometry is great.

Having done supernova photometry for over 15 years, I can assure you that my
local standards agree to 0\fm02 and that the photometry is great. But there
are some subtle problems that are not usually revealed. In Figure \ref{fig2} I
show the data for the Type Ia supernova SN 1998bu as published in
\cite{Sunetal99} based on CTIO 0.9m CCD data. Included in this diagram are
photometric points from the CTIO YALO 1m telescope. The YALO data were reduced
using color terms appropriate for that telescope/detector and the same local
standards as used for the reductions of the data from the 0.9m telescope.
Said another way, {\it the two data sets come from the facility instruments on
two telescopes side-by-side on CTIO and reduced using the same local standards
by me.} The YALO data did not start until 35 days after maximum light.

Plotted at the typical scale of a supernova light curve, the data seem to be
similar, but actually they are quite different. The differences between the
data sets are: $\delta(U,B,V,R,I)=(-0.41,-0.7,0.06,0.00,0.05)$ in the sense of
(0.9m - YALO). The color difference is $\delta(B-V) = -0.12$. The large
difference in the $U$ photometry is not surprising. The 0.9m data were taken
with a Tek (now SITe) CCD which has no response shortward of 3500A whereas the
YALO telescope CCD is a Loral device which has excellent ultraviolet response.
In general, most CCD $U$ photometry for supernovae is suspect at the 0\fm1
level, unless it was taken with a CCD with a flat quantum efficiency in $UB$,
such as coated GEC, EEV, or TI chips.

The large difference in the $B-V$ colors can be easily reproduced by
performing synthetic photometry using the actual filter/detector transmission
curves convolved with spectrophotometry of standard stars and the
supernova. These errors come from the fact that the supernova flux
distribution is very different than the defining stellar fluxes of the
standard stars. Near maximum light where a supernova spectrum is nearly
stellar (except in $U-B$), the differences between data sets are very small,
with $\delta(B-V) \leq 0.04$. Similar differences in photometry have been
reported for SN 1987A (\cite{Sunetal88,Men89})

However, the point to be made here is neither the 0.9m nor the YALO
transmissivities are the correct transmissivities as defined by Bessell
(\cite{Bes90}) for the Kron-Cousins $UBVRI$ photometric system. The
transformed supernova photometry will have systematic errors even though we
have performed the usual color and extinction corrections. In the future,
photometrists will need to correct their photometry not only to the standard
Kron-Cousins $UBVRI$ as done now, but also correct the photometry to a
standard filter transmissivity curve using spectrophotometry, in a manner
identical to K-corrections. The C\'alan/Tololo group has made a large atlas of
spectrophotometry of supernova spectra and Landolt standard stars which will
be made available to the community to perform these corrections.

\subsection{Other Topics}

\subsubsection{What is Causing the Secondary Maximum?}

The secondary maximum at about 30 days past blue maximum is a prominent
feature in Type Ia light curves. It is seen in $IJHK$ and the precise date of
secondary maximum is not a constant, but is roughly a function of \dmm.  It
gets much smaller for the low luminosity supernovae and is missing in SN
1991bg.  A small inflection can be seen in the $R$ light curve, and even in
the $V$ curve. The secondary peak has also been seen in the $I$ band light
curve of a supernova at $z \approx 0.5$ (\cite{Rieetal00}). The ``uvoir''
bolometric light curves also include this inflection point
(\cite{Sun96,Conetal00}) which seems to indicate that flux has been
redistributed from the blue to the red. Due to its presence across many
photometric bandpasses, it is obvious this feature is not a spectral feature
but is likely due to the time evolution of the line opacities
(\cite{Spyetal94,Wheetal98}). Since the peak is so prominent and a function of
luminosity, it is important that the precise physical reason for its
variability be established.


\subsubsection{Where is the Hydrogen?}

One thing we can all agree upon at this conference is that Type Ia supernovae
are characterized by a lack of hydrogen in the observed spectrum. This is
curious because most of the models require a companion star to dump material
onto the white dwarf to bring it to a critical mass for explosion. It is
difficult to imagine such mass transfer without large amounts of hydrogen
being present, unless the explosion is due to the merger of a double
degenerate white dwarf pair. The DD scenario is less attractive at this time
due to the observed lack of DDs that are close enough to merge in a Hubble
time. As discussed in \cite{Chu86,Whe92,Maretal00}, the interaction of the
supernova ejecta and the secondary is expected to produce 0.1-0.5\solarm\ at
velocities below 1000 \kms, depending on the nature (dwarf/giant) of the
progenitor, with a small high-velocity tail. This hydrogen (and smaller
amounts of helium) will be present a few months after maximum mixed in with
the low velocity iron layer and may be detectable with high S/N
high-dispersion spectra. A number of attempts have been made to find this
hydrogen and helium, but there has been no convincing evidence found yet,
partly because of the confusion in the spectra, and partly due to the lack of
high quality data on Type Ia supernovae. This is a project that needs to be
done right now with an echelle spectrograph on an 8-10m class telescope.

It is important not only to observe Type Ia's a few months after maximum to
search for hydrogen, but also at the time of maximum. It is possible that
transient narrow H$_\alpha$\ may be present from a relic wind from the
secondary (\cite{Whe92}), either in emission or absorption. 

\subsubsection{Velocity Structure at Maximum Light}

Nugent et al (\cite{Nugetal95}) have shown that the spectral diversity of Type
Ia supernovae at maximum light can be explained by a small range in \teff\
from 7500-11000K. The close relationship between the spectral features and the
luminosity (as measured by \dmm\ or its equivalent) has been used to create
spectral methods for correcting peak luminosity into a standard value.
However, while the spectral features may be closely related to \dmm\, the
velocity at maximum light is not clearly related to \dmm. Wells et al
(\cite{Weletal94}) have shown that the velocity at maximum light for the SiII
line at 6355\AA\ is basically uncorrelated with the intrinsic luminosity.  The
CaII lines show a weak correlation with \dmm. It has been claimed that the
velocity is correlated with galaxy type, with the early-type galaxies showing
lower peak velocities (\cite{Bravan93}). A large number of supernovae
(M. Phillips, private communication) does not show this trend.

It is rather worrisome that we are presently using a single physical parameter
such as \dmm\ to correct to a standard luminosity, while another fundamental
parameter, velocity, which related to the kinetic energy of expansion and
therefore the overall energy budget of the explosion, seems to be uncorrelated
with standard luminosity. What, then, are the line velocities telling us about
the explosion? This is an important question where the observers need some
guidance from theory.


\smallskip\smallskip\smallskip\smallskip

I would like to thank Mario Hamuy, Mark Phillips, Phillip Pinto, Brian
Schimdt, and Robert Schommer for our many discussions about supernovae. If
there are errors in this paper, it can only be my fault.



\end{document}